# NaIrO$_3$ – A pentavalent post-perovskite

M. Bremholm[a]*, S.E. Dutton[a], P.W. Stephens[b] and R.J. Cava[a]

[a] Department of Chemistry, Princeton University, Princeton, NJ 08544, USA
[b] Department of Physics and Astronomy, Stony Brook University, Stony Brook, NY 11794, USA

* Corresponding author: bremholm@princeton.edu

**Abstract**

Sodium iridium (V) oxide, NaIrO$_3$, was synthesized by a high pressure solid state method and recovered to ambient conditions. It is found to be isostructural with CaIrO$_3$, the much-studied structural analog of the high-pressure post-perovskite phase of MgSiO$_3$. Among the oxide post-perovskites, NaIrO$_3$ is the first example with a pentavalent cation. The structure consists of layers of corner- and edge-sharing IrO$_6$ octahedra separated by layers of NaO$_8$ bicapped trigonal prisms. NaIrO$_3$ shows no magnetic ordering and resistivity measurements show non-metallic behavior. The crystal structure, electrical and magnetic properties are discussed and compared to known post-perovskites and pentavalent perovskite metal oxides.

**Research highlights:** ► NaIrO$_3$ post-perovskite stabilized by pressure. ► First example of a pentavalent oxide post-perovskite. ► Non-metallic and non-magnetic behavior of NaIrO$_3$.

**Introduction**

MgSiO$_3$ is a major component of the Earth's lower mantle and its structural transformations under pressure are of significant importance in the geosciences. In 2004 Murakami *et al*. [1] and Oganov and Ono [2] independently showed that MgSiO$_3$ transforms from a distorted perovskite to the high-pressure CaIrO$_3$ structure-type at a pressure of ~120 GPa. Until 2004 CaIrO$_3$ was the only oxide known to crystallize in this structure type, which is now known as the post-perovskite structure. However, since 2004 several isostructural oxide post-perovskites have been identified: CaPtO$_3$ [3,4], CaRuO$_3$ [5], CaRhO$_3$ [6], CaSnO$_3$ [7], MgGeO$_3$ [8], and MnGeO$_3$ [9]. If including non-oxides, approximately 20 additional members are known, *e.g.* NaMgF$_3$ [10], NaZnF$_3$ [11], ThMnSe$_3$ [12], and UMnSe$_3$ [12]. The physical properties of post-perovskites have been widely investigated as model systems for the post-perovskite MgSiO$_3$ [13]. This family of structures is also of interest in materials science and, for example, Ohgushi *et al*. [14] found that hole-doping of CaIrO$_3$ leads to a metal-insulator transition (MIT) concurrent with a destabilization of the long-range antiferromagnetic order.

High pressure solid state synthesis with solid oxidants provides the means for stabilization of metals in high oxidation states with correspondingly smaller ionic radii, and consequently can lead to the preparation of new compounds and structure types. Only a limited number of Na-Ir-O compounds are known, *i.e.* Na$_4$IrO$_4$ [15], Na$_2$IrO$_3$ [16], and Na$_4$Ir$_3$O$_8$ [17], which all contain Ir(IV). Iridate structures are mostly found with tetravalent iridium but some pentavalent compounds are known, *e.g.* KIrO$_3$ [18], Ba$_{0.5}$IrO$_3$ [19], La$_2$LiIrO$_6$ [20], and Ln$_3$IrO$_7$ (Ln = Pr, Nd, Sm, Eu) [21] as well a small number of hexavalent compounds, *e.g.* Nd$_2$K$_2$IrO$_7$ [22] and Sr$_2$CaIrO$_6$ [23]. In the search for a novel pentavalent iridate perovskite we found that for NaIrO$_3$ the post-perovskite structure can be stabilized by pressure, the first example of a pentavalent cation in the post-perovskite structure type. Here, we report the synthesis and crystal structure of NaIrO$_3$ and discuss the crystal chemistry. Furthermore, we have measured the physical properties and discuss these in relation to expectations based on band structure calculations.

**Materials and methods**

NaIrO$_3$ was prepared from a 1.5:1:4 ratio of Na$_2$O$_2$ (95%, Alfa Aesar), Ir (99.95%, Alfa Aesar), and NaClO$_3$ (99%, Alfa Aesar), where the latter serves only as an oxidizer. The Na$_2$O$_2$ excess of 50% was used to compensate for loss of sodium during the reaction. The precursor powders were weighed, ground and loaded into a gold capsule in an argon glove box. The gold capsule was placed in a boron nitride crucible and inserted in a pyrophyllite cube with a resistive graphite furnace. The sample was pressurized to 4.5 GPa in a cubic multi-anvil press (Rockland Research Corp.). The temperature was increased to 800°C at a rate of 50°C/min. The temperature was controlled by an internal thermocouple and kept constant for 30 min and then cooled at 50°C/min before the pressure was released. The sample was washed with deionized water to remove the NaCl produced by the decomposition of the sodium chlorate oxidant and then washed with methanol and dried at room temperature.

The polycrystalline sample was characterized by synchrotron powder X-ray diffraction (PXRD) at beamline X16C at NSLS using a 0.3 mm glass capillary ($\lambda$ = 0.69910 Å). A linear absorption coefficient, $\mu$ = 2.9 cm$^{-1}$, was calculated from the elemental composition, X-ray attenuation coefficients, and an estimated capillary packing density. The diffraction peak profile was modeled using a Thompson-Cox-Hasting pseudo-Voigt function and the background was modeled using an 8th order polynomial. The magnetic susceptibility was measured from 2 to 300 K in a magnetic field of 1 T, the electrical resistivity from 2 to 300 K, and the specific heat from 0.4 to 20 K using a compressed powder sample (Quantum Design Physical Property Measurement System, PPMS). The resistivity measurement was performed on a polycrystalline pellet (pressed at 4.5 GPa) using the standard four-probe method with platinum leads and silver paste.

Electronic band structure calculations of NaIrO$_3$ were performed using the local density approximation (LDA) of density functional theory (DFT) as implemented in the WIEN2k package [24], which is based on a full-potential linear augmented-plane-wave method (FP-LAPW). The experimental lattice parameters and atomic coordinates were used (Table 1) with atomic sphere radii of 2.31, 1.89, and 1.67 a.u. for Na, Ir, and O, respectively. The Perdew-Burke-Ernzerhof generalized gradient approximations potentials (PBE-GGA) were used. The energy separating core and valence states was set to -7.2 Ry to avoid core charge leaking. The cut-off wave-number $K_{max}$ for wave functions in interstitial regions was set such that $R_{mt}K_{max}$ = 7, where $R_{mt}$ is the smallest atomic sphere radius. The integration over Brillouin zone was performed using 2000 k-points (247 k-points in the irreducible zone). The energy convergence criterion was 10$^{-5}$ eV. No changes were observed upon lowering convergence criterion or increasing the number of k-points or $R_{mt}K_{max}$. Spin-orbit coupling was included in the calculation for the iridium atom as well as the relativistic local orbital basis function ($p_{\frac{1}{2}}$). For LDA+$U$ calculations the $U_{eff}$ ($U_{eff} = U - J$) was set to 1 eV, which is the approximate value expected for iridium [25].

**Results**

The powder X-ray diffraction pattern was initially indexed to an orthorhombic unit cell, $a$ = 3.04 Å, $b$ = 10.36 Å, $c$ = 3.59 Å (DICVOL04 program [26]). However, closer inspection revealed a weak peak from a 2$c$ super cell (($hkl$) = (131)) and extinction rules were used to identify a $C$-centered unit cell and the structural solution verified the space group as *Cmcm* (#63). The structure was solved using direct methods (FOX program [27]) and the atomic coordinates were refined using the Rietveld method (FullProf program suite [28]). Fig. 1 shows the excellent quality of the fit. Refinement of the sodium site occupancy results in full occupancy within the uncertainty and the occupancy was fixed to the stoichiometric value for the final refinement. The crystallographic data are listed in Table 1. Ir metal is present in all our preparations of NaIrO$_3$, it is a common impurity in the synthesis of Ir oxides due to the stability of metallic Ir in oxidizing ambient at high temperatures. Its content varies from approximately 5 to 20 wt% in different preparations and different positions in pressure cell. Physical property measurements were made on samples that had 10% or less Ir metal impurity.

The pentavalent iridium atoms in post-perovskite NaIrO$_3$ are six-fold coordinated to oxygen ions in octahedra. These octahedra share edges along [100] and corners along the [001], *i.e.* there are sheets of octahedra with Ir atoms in a tetragonal arrangement in the *ac*-plane (Fig. 1A, inset). The average Ir-O bond length is 1.898 Å with little difference in the individual Ir-O bond lengths, 0.0138 Å (2 x 1.9068(4) and 4 x 1.8930(16) Å, Fig. 1B). However, the shape of the octahedra is strongly deformed, showing an O2-Ir-O2 angle of 73.2° instead of the 90° in a regular octahedron (Fig. 1C). The Ir-Ir distance between edge sharing octahedra is 3.03967(3) Å, which means that there is no metal-metal bonding because Ir-Ir bond distances typically fall in the range 2.6-3.0 Å [29]. On the contrary, the strong deformation of the octahedra indicates a strong electrostatic repulsion of the pentavalent iridium ions. The sodium atoms are located in between the layers and are coordinated to 8 oxygen atoms with an average bond length of 2.572 Å. The NaO$_8$ polyhedron is a bicapped trigonal prism (Fig. 1D).

**Physical properties**

The magnetic susceptibility of NaIrO$_3$ shows no evidence for magnetic ordering below 300 K and follows Curie-Weiss behavior in addition to a dominant negative temperature independent contribution over much of the temperature range of measurement (Fig. 3). Thus, the observed magnetic susceptibility was modeled by:

$$\chi(T) = C/(T-\theta_W) + \chi_0 \qquad (1)$$

where $C$ is the Curie constant, $\theta_W$ is the Weiss temperature and $\chi_0$ is the temperature-independent diamagnetic term. The Curie-Weiss parameters obtained are: $\theta_W$ = -2.2 K, $C$ = 9.7 x 10$^{-3}$ emu K mol$^{-1}$ Oe$^{-1}$, $\mu_{eff}$ = 0.28 $\mu_B$/Ir, and $\chi_0$ = -1.9 x 10$^{-3}$ emu mol$^{-1}$ Oe$^{-1}$ (corrected for the elemental core diamagnetism of Ir$^{5+}$ (-2 x 10$^{-5}$ emu mol$^{-1}$ Oe$^{-1}$ [30]) and the sample container contribution (-2.5 x 10$^{-4}$ emu mol$^{-1}$ Oe$^{-1}$)). The observed weak moment of $\mu_{eff}$ = 0.28 $\mu_B$/Ir is much smaller than the expected moment for a low spin 5$d^4$ ion in an octahedral field with degenerate $t_{2g}$ orbitals ($t_{2g}^4$, $S$ = 1, $\mu_{eff}$ = 2.83 $\mu_B$/Ir). This may represent a low intrinsic itinerant moment, but given the low value of $\theta_W$ and the small observed Curie constant, it is likely that the low temperature Curie tail is due to the presence of a small proportion of paramagnetic impurities in the sample: 0.9 mole % of a spin 1 impurity would account for the paramagnetism.

The electrical resistivity at room temperature is 0.06 Ω cm and increases monotonically upon cooling, reaching 400 Ω cm at 2 K (Fig. 3). These data indicate that NaIrO$_3$ is non-metallic. An effective activation energy electron conduction model, as would be seen for a simple semiconductor, predicts an exponential temperature dependent resistivity, $\rho(T) \propto \exp(\Delta E_{eff}/k_BT)$, i.e. a linear behavior of log($\rho$) vs. 1/$T$ (Fig. 3, inset A). No linear temperature range is found. Instead a variable range hopping (VRH) model [31], $\rho(T) = \rho_0\exp(T_0/T^{1/(d+1)})$, where $d$ is the dimensionality of the hopping, is found to describe the resistivity well for $T$ < 50 K with a dimensionality, $d$ = 2 and $T_0$ = 1.7(1) x 10$^3$ K. This model is inadequate for higher temperatures, however, where no simple model fits the observed data. The contribution from grain-boundary resistance is expected to be insignificant because the pellet was pressed at 4.5 GPa. The only impurity detectable with PXRD is metallic Ir. Although the influences of grain boundary resistivity and Ir metal impurity may be present, we conclude that NaIrO$_3$ has an unusual hopping-type resistivity, from which we infer the presence of a defective lattice and a low carrier concentration in a material that is intrinsically a semiconductor. Further experiments on single crystals will be of interest for a detailed analysis.

The low temperature heat capacity ($0.4 < T < 20$ K) shows no peaks, consistent with the absence of low temperature magnetic ordering or structural phase transitions. A plot of $C_p/T$ vs. $T^2$ shows linear behavior from 3 K to 10 K (Fig. 4). Below 3 K, $C_P/T$ diverges as $T \rightarrow 0$ K, and the data was fitted by including an additional term:

$$C_p = A + \gamma T + \beta T^3 \qquad (2)$$

where the $\gamma$ and $\beta$ are the coefficients for the electronic and the lattice vibration contributions to the specific heat, respectively. The additional term, $A$, describes a constant contribution that only becomes significant below 3 K ($A = 5.9(2)$ mJ mol$^{-1}$ K$^{-1}$). The origin of this contribution is not known. The value of $\beta$ is $0.3110(4)$ mJ mol$^{-1}$ K$^{-4}$ and corresponds to a Debye temperature, $\theta_D$, of $315(1)$ K. The electronic contribution to the specific heat is found to be small, $\gamma = 4.0(2)$ mJ mol$^{-1}$ K$^{-2}$. Although the measurements show that NaIrO$_3$ is highly resistive at low temperatures, the small but non-zero electronic specific heat coefficient, $\gamma$, suggests that the electronic structure of NaIrO$_3$ may not be fully gapped at the Fermi level.

**Electronic structure calculation**

To further illuminate the electronic properties of NaIrO$_3$ we calculated the electronic band structure. The rather weak orbital-dependence of the exchange correlation energy in DFT calculations means that the strong on-site Coulomb repulsion in systems with narrow $d$-bands is underestimated. Consequently, the DFT calculations often produce metallic ground states or band gaps smaller than experimentally observed. LDA+$U$ corrections are commonly used to obtain a more accurate description. Thus, we applied the LDA+$U$ corrections to the 5$d$-wave functions of NaIrO$_3$. Furthermore, spin-orbit coupling can have a strong influence on the electronic band structures for heavy elements such as Ir, and is exemplified by the electronic structure of Sr$_2$IrO$_4$ [32]. The electronic band structure is shown in Fig. 5. In spite of the inclusion of SO coupling and a moderate $U_{\text{eff}}$ value, the band structure shows a significant DOS at the Fermi level, $n(E_F)$, indicating that NaIrO$_3$ should be metallic, contradicting the experimental resistivity data. For comparison we also calculated the electronic structure using

LDA. When comparing a simple non-magnetic calculation and the spin-polarized LDA+$U$+SO there are only very small differences and the general characteristics of the density of state (DOS) are reproduced (Fig. 5 B and C).

The 5$d$ orbitals are more extended compared to 4$d$ and particularly the 3$d$ orbitals. Hybridization of the Ir 5$d$ and O 2$p$ orbitals is therefore much more pronounced, as are the 5$d$-5$d$ orbital interactions. The LDA band structure features the Ir 5$d$ bands of $e_g$ and $t_{2g}$ components separated by a large octahedral crystal field splitting, $\Delta_{oct}$ = 2 eV, which, for NaIrO$_3$ is extends from ~0.9 to ~3 eV above $E_F$. The $e_g$ bands are located 3 eV above $E_F$ and the $t_{2g}$ bands extend from 2 eV below $E_F$ to 0.9 eV above $E_F$. A nominal configuration of 5$d^4$ Ir$^{5+}$ may reasonably be expected due to simple electron counting and the octahedral crystal field, forming paired electrons in the 5$d_{zx}$ and 5$d_{yz}$ orbitals, resulting in a non-magnetic ground state of $S$ = 0. However, the calculations indicate that this simple orbital picture does not hold. Although the transport properties definitively show NaIrO$_3$ to be non-metallic, the electronic structure calculations, even including a reasonable value for $U_{eff}$ and spin-orbit coupling, indicate that NaIrO$_3$ should be metallic. The reasons for the problem with the modeling of the system by electronic structure calculations are not known, but it may be that either $U_{eff}$ is much larger than the value we have employed, or the compound is antiferromagnetically ordered at room temperature; such ordering could open up a gap within the $t_{2g}$ 5$d$ bands.

Similar band structure characteristics have been reported for the post-perovskite CaRhO$_3$ (Rh$^{4+}$, 4$d^5$) [33,34]. As for NaIrO$_3$, LDA+$U$ calculations suggest that CaRhO$_3$ should be metallic, while the experimental resistivity shows that it is a thermally activated semiconductor with $\Delta E_{eff}$ = 38 meV. However, unlike NaIrO$_3$, CaRhO$_3$ shows evidence for unusual AFM ordering at 90 K with 2.99 $\mu_B$/Rh and $\theta_W$ = -1071 K [34]. Other post-perovskites such as CaPtO$_3$ [3] and CaIrO$_3$ [35] also show semiconducting or insulating behavior, and CaRuO$_3$ (Ru$^{4+}$, 4$d^4$) shows low-dimensional magnetism at 470 K, and long-range magnetic order near 270 K [36].

**Crystal chemistry of post-perovskites**

Comparing the unit cell of NaIrO$_3$ to those of the known CaB$^{4+}$O$_3$ oxide post-perovskites (B = Ir, Pt, Ru, Rh), shows that the $a$ and $c$ dimensions of NaIrO$_3$ are shorter than those of CaB$^{4+}$O$_3$, while the $b$ axis is longer in NaIrO$_3$ (Table 2). This is as expected based on ionic radii, $r_i(Ir^{5+}) < r_i(B^{4+})$ (57 pm vs. 62-62.5 pm) and $r_i(Na^+) > (Ca^{2+})$ (118 pm vs. 112 pm) [37]. When comparing the bond angles, NaIrO$_3$ shows both the largest octahedral tilting and largest deformation of the BO$_6$ octahedra.

In NaIrO$_3$ the bond valence sum (BVS) of the Na$^+$-O is 1.0 as expected, while that of Ir$^{5+}$-O is 6.3 [38], much larger than expected for Ir$^{5+}$. The over-bonding can to some extent be explained by the strong deformation of the octahedra and the expected high covalence of the Ir$^{5+}$-O bond. However, for other pentavalent iridium oxides, the average Ir-O distances are typically in the range 1.92 to 1.99 Å [18,21], significantly longer than the 1.90 Å reported here for NaIrO$_3$. Bond valence calculations are referenced to average values and may not be applicable in all cases. The NaIrO$_3$ post-perovskite displays a different type of structure and no other Ir(V)-O bonds are available for a direct comparison. The apparently short Ir-O bond lengths in NaIrO$_3$ could be an underestimate caused by the presence of correlated atomic motion, just as has been observed in the CaIrO$_3$ post-perovskite by comparison of Rietveld refinement and total scattering modeling (PDF), which found that the Rietveld determined Ir$^{4+}$-O bond lengths were underestimated by ~0.15 Å at 300 K [39]. Alternatively, the strong deformation of the octahedra could be a factor, indicative of strong crystallographic strain in the system, which significantly impacts bond valence determinations.

Table 3 summarizes the bonding geometry for NaIrO$_3$ and previously reported post-perovskite metal oxides. In all cases of post-perovskite oxides the B-O bonds have strong covalent character. The largest B-O bond length differences are observed for CaIrO$_3$ (Ir$^{4+}$, $5d^5$, $t_{2g}^5$) and CaRuO$_3$ (Ru$^{4+}$, $4d^4$, $t_{2g}^4$) and must be attributed to Jahn-Teller distortion, which is not expected for CaPtO$_3$ (Pt$^{4+}$, $5d^6$, $t_{2g}^6$). Although NaIrO$_3$ (Ir$^{5+}$, $5d^4$, $t_{2g}^4$) is weakly Jahn-Teller active, it shows the lowest bond difference and is the only compound in the family for which the metal-oxygen apex bond is the shorter bond (although not by much). All oxide post-perovskites have a high electronegativity of the B ions, (*e.g.* $\chi_{P,Ir}$ = 2.20, $\Delta\chi_P$(Ir-O) = 1.24, covalent character $f_i = \exp(-\Delta\chi_P^2/4) = 0.7$, Table 5).

The best-known measure of perovskite distortion is the Goldsmith tolerance factor, $t = (r_A + r_X)/(\sqrt{2}(r_B + r_X))$, where $r_A$, $r_B$ and $r_X$ are tabulated values [37] of the ionic radii at ambient pressure of the A, B and X ions, respectively. For ABX$_3$-type post-perovskites, Ohgushi *et al.* introduced an extension of the tolerance factor $t' = d_{A-X}/(\sqrt{2}d_{B-X})$, where $d_{A-X}$ and $d_{B-X}$ are the experimentally determined average A-X and B-X bond distances, respectively [3]. Table 5 shows the perovskite and post-perovskite tolerance factors for most of the known $ABX_3$-type post-perovskites.

The tolerance factors of previously reported post-perovskites all fall in the ranges from $0.75 < t < 0.90$ and $0.78 < t' < 0.87$ with an average value of 0.82 in both cases. Post-perovskites with chalcogenides show the lowest values ($0.752 < t < 0.78$), fluorine and iodine compounds show intermediate values ($0.81 < t < 0.87$) while the oxides show the highest values ($0.84 < t < 0.90$). The case of $NaIrO_3$ stands out with the highest values of $t$ and $t'$ reported so far (~0.93 and ~0.96, respectively). The values of both tolerance factors ($t$ and $t'$) for $NaIrO_3$ fall comfortably within the range observed for perovskites, $0.75 < t < 1.05$, and thus clearly the expected orthorhombic distortion suggested by the tolerance factor alone has a low predictive power for the existence of post-perovskites.

Thus, despite the fact that $NaIrO_3$ has a nearly ideal Goldsmith perovskite tolerance factor, high pressure stabilizes the post-perovskite rather than the perovskite phase. This is likely due to the covalence of the bonding, which causes a linear O-Ir-O geometry to be unfavorable. The covalence of the Ir-O bond as a controlling factor in the stabilization of the post-perovskite has been previously discussed by Ohgushi *et al* [3].

An investigation of the structural transformation paths deepens the understanding of the driving force that leads to stabilization of the post-perovskite structure and gives important insight useful in the search for new post-perovskite compounds. As in $MgSiO_3$, most post-perovskites transform under pressure from the $GdFeO_3$-type perovskite (S.G. *Pnma*) to the post-perovskite. Approximately 60% of perovskites are of the $GdFeO_3$-type [40] and a high number of post-perovskites are therefore expected to be stabilized under pressure, although relatively few have been observed. In at least one case, $CaRhO_3$, an intermediate monoclinic phase was observed before the transformation to post-perovskite.

The ratio of the volume of the $AO_{12}$ and $BO_6$ polyhedra in $GdFeO_3$-type perovskites that transform to post-perovskites shows a trend identified by Martin *et al.* [41] The $AO_{12}$ to $BO_6$ polyhedra volume ratio can be expressed as $V_A/V_B = (V_{Cell}/4 - V_B)/V_B$. In the post-perovskites, the space not occupied by $BO_6$ octahedra is assigned to $AO_8$. For all cases where the post-perovskite is stable under ambient conditions, $V_A/V_B < {\sim}4$ (Table 4). For systems where the polyhedra ratio decreases with pressure, the post-perovskite can be pressure-stabilized, *i.e.* transitions can occur when $dV_A/dP > dV_B/dP$ and are found to occur approximately when $V_A/V_B < {\sim}4$ (Table 4).

Another strong indicator of the potential for a phase transformation between perovskite and post-perovskite forms is the octahedral tilt-angle of the GdFeO$_3$-type perovskites, $\Phi = \cos^{-1}(\sqrt{2}a^2/bc)$. Transformation to the post-perovskites occurs when the $\Phi$ reaches a value of 25°, which is effectively a different expression of the condition $V_A/V_B \cong 4$. For cases where $\Phi$ in the perovskite is larger than 13° at ambient conditions, the post-perovskite is often found to be quenchable. In cases where $\Phi$ is smaller, no transformations are observed since the perovskites become more regular. The study by Tateno *et al.* shows this effect: under pressure MnGeO$_3$ becomes increasingly distorted while the opposite occurs for CdGeO$_3$ [7,9].

The present study expands the chemical variation of post-perovskite oxides to include pentavalent metal ions. The large polyhedra volume ratio of NaIrO$_3$ ($V_A/V_B = 5.48$) compared to tetravalent post-perovskites suggests that the transition to post-perovskite for pentavalent cases can occur at higher ratios. Additional compounds with similar chemistry can be expected to be stabilized in this structure type under pressure as well. These have to be highly electronegative M ions, making the most obvious candidates the 4$d$ and 5$d$ transition metals, and, in particular, the platinum group metals. Mo, W, and Re, although being highly electronegative, form cubic bronzes under pressure. Nb and Ta form distorted perovskites but the electronegativities of these ($\chi_{P,Nb} = 1.6$ and $\chi_{P,Ta} = 1.5$) are too low to stabilize the post-perovskite structure except potentially at extreme pressures as for MgSiO$_3$ ($\Delta\chi_P$(Si-O) = 1.54) and unquenchable to ambient pressure. Of particular interest would be the preparation of quenchable ABO$_3$ post-perovskites where the B cations have unpaired electrons and anisotropic magnetic properties and electron-electron correlations are expected. Currently, only three such systems are known (CaRuO$_3$, CaRhO$_3$, and CaIrO$_3$).

**Conclusion**

A novel post-perovskite compound, NaIrO$_3$, with pentavalent iridium ions was obtained by high-pressure synthesis at a relatively moderate pressure of 4.5 GPa. The crystal structure refinement shows that NaIrO$_3$ has substantially deformed IrO$_6$ octahedra with rather short Ir-O bonds, consistent with the covalence of the bonds and electrostatic repulsion of the Ir atoms. Measurement of the resistivity shows a non-trivial behavior, fitting variable range hopping in 2D at temperatures below 50 K. The magnetic susceptibility shows only a weak moment, suggesting that either the Ir is non-magnetic in this compound or that magnetic ordering may have occurred above ambient temperature. Band structure calculations that include spin-orbit coupling and a moderate value of electron repulsion do not explain the observed highly resistive character of the phase. The discovery of a $NaIrO_3$ post-perovskite expands the chemical variation of the post-perovskite family and suggests that a significant number of compounds with this structure type are still to be found, mainly within the platinum group metals.

**Acknowledgments**

Funding for this research was provided primarily by the NSF Solid State Chemistry program, grant DMR 0703095. Use of the National Synchrotron Light Source, Brookhaven National Laboratory, was supported by the U.S. Department of Energy, Office of Science, Office of Basic Energy Sciences, under Contract No. DE-AC02-98CH10886. M. Bremholm gratefully acknowledges funding from the Villum Kann Rasmussen foundation.

# Tables

Table 1. Unit cell dimensions, atomic coordinates and isotropic thermal displacement parameters of NaIrO$_3$ at ambient conditions, space group *Cmcm* (#63).

| $a = 3.03968(3)$ Å | $b = 10.3576(12)$ Å | $c = 7.1766(3)$ Å | $V = 225.97(1)$ Å$^3$ |
|---|---|---|---|
| $R_B = 0.065$ | $R_F = 0.050$ | $R_{wn} = 0.20$ | $\rho = 7.737$ g cm$^{-3}$ |

| Atom | Site (Wyckoff) | x | y | z | $B_{iso}$ (Å$^2$) |
|---|---|---|---|---|---|
| Na1 | 4c | 0 | 0.2507(8) | ¼ | 1.8(2) |
| Ir1 | 4a | 0 | 0 | 0 | 0.39(3) |
| O1 | 4c | ½ | 0.43766(10) | ¼ | 0.35(18)* |
| O2 | 8f | ½ | 0.10159(6) | 0.0568(10) | 0.35(18)* |

* Isotropic thermal parameters for the two oxygen sites were constrained to the same value.

Table 2. Crystallographic data for NaIrO$_3$, previously reported quenchable post-perovskite metal oxides and MgSiO$_3$. CaSnO$_3$ is not included because only the unit cell volume at ambient pressure was reported by Tateno *et al.* [7])

| ABO$_3$ | | | NaIrO$_3$ | CaIrO$_3$ | CaPtO$_3$ | CaRuO$_3$ | CaRhO$_3$ | MgSiO$_3$* |
|---|---|---|---|---|---|---|---|---|
| Ref. | | | *present study* | Hirai [42] | Ohgushi [3] | Kojitani [5] | Shirako [6] | Murakami [1] |
| a [Å] | | | 3.03968(3) | 3.1430 | 3.12607 | 3.1150 | 3.1013 | 2.456 |
| b [Å] | | | 10.3576(12) | 9.8569 | 9.91983 | 9.8268 | 9.8555 | 8.042 |
| c [Å] | | | 7.1766(3) | 7.2924 | 7.35059 | 7.2963 | 7.2643 | 6.093 |
| V [Å$^3$] | | | 225.95(3) | 225.92 | 227.942 | 223.34 | 222.03 | 120.34 |
| A 4c | x | | 0 | 0 | 0 | 0 | 0 | 0 |
| | y | | 0.2507(8) | 0.25025 | 0.2500 | 0.2512 | 0.2521 | 0.253 |
| | z | | ¼ | ¼ | ¼ | ¼ | ¼ | ¼ |
| B 4a | x | | 0 | 0 | 0 | 0 | 0 | 0 |
| | y | | 0 | 0 | 0 | 0 | 0 | 0 |
| | z | | 0 | 0 | 0 | 0 | 0 | 0 |
| O1 4c | x | | ½ | ½ | ½ | ½ | ½ | ½ |
| | y | | 0.43766(10) | 0.4243 | 0.4253 | 0.4308 | 0.4279 | 0.423 |
| | z | | ¼ | ¼ | ¼ | ¼ | ¼ | ¼ |
| O2 8f | x | | ½ | ½ | ½ | ½ | ½ | ½ |
| | y | | 0.10159(6) | 0.1276 | 0.1230 | 0.1281 | 0.1226 | 0.131 |
| | z | | 0.05681(10) | 0.0507 | 0.0451 | 0.0528 | 0.0548 | 0.064 |

* 121 GPa-300 K.

Table 3. Bonding geometry for NaIrO$_3$ and previously reported post-perovskite metal oxides, ABO$_3$.

|  | NaIrO$_3$ | CaIrO$_3$ | CaPtO$_3$ | CaRuO$_3$ | CaRhO$_3$ | MgSiO$_3$* |
|---|---|---|---|---|---|---|
| Ref. | *present study* | Hirai [42] | Ohgushi [3] | Kojitani [5] | Shirako [6] | Murakami [1] |
| Interatomic distance [Å] | | | | | | |
| B-O1  x 2 | 1.9068(4) | 1.970 | 1.981 | 1.947 | 1.950 | 1.644 |
| B-O2  x 4 | 1.8930(16) | 2.047 | 2.010 | 2.039 | 2.006 | 1.664 |
| Average B-O | 1.8976 | 2.021 | 2.000 | 2.008 | 1.987 | 1.657 |
| B-O1 - B-O2 | 0.0138 | -0.026 | -0.010 | -0.031 | -0.056 | -0.020 |
| B-site BVS | 6.3 | 4.0 | 4.3 | 3.8 | 4.0** | 5.5 |
| A-O1  x2 | 2.461(7) | 2.330 | 2.3382 | 2.353 | 2.325 | 1.838 |
| A-O2  x4 | 2.573(6) | 2.454 | 2.5085 | 2.441 | 2.458 | 1.938 |
| A-O2  x2 | 2.681(8) | 2.504 | 2.5097 | 2.508 | 2.535 | 2.129 |
| Average | 2.572 | 2.436 | 2.4662 | 2.436 | 2.444 | 1.961 |
| Bond angle [°] | | | | | | |
| O1-B-O2 | 90.826(6) | 93.76 | 94.25 | 92.2 | 92.0 | 91.2 |
| O2-B-O2 | 73.186(1) | 79.66 | 77.94 | 80.4 | 78.7 | 84.9 |
| B-O1-B | 140.414(1) | 135.47 | 136.08 | 139.0 | 137.3 | 135.8 |

* 121 GPa, 300 K

** No $R_0$ value has been reported for Rh$^{4+}$-O and the value for Ru$^{4+}$-O has been assumed.

Table 4. Tilt angle, Φ, for perovskites (PV, GdFeO$_3$-type) and polyhedra ratio, $V_A/V_B$, for the corresponding post-perovskites.

| Compound | PV, octahedral tilt, Φ = cos$^{-1}$($\sqrt{2}a^2/bc$) | Polyhedra ratio ($V_{Cell}/4 - V_B$)/$V_B$ | Reference |
|---|---|---|---|
| NaIrO$_3$ | no perovskite known | 5.48 | *present study* |
| MgSiO$_3$* | Unstable at 1 atm. | 3.97 | Murakami *et al.* 2004 [1] |
| MgGeO$_3$* | Unstable at 1 atm. | 4.28 | Hirose *et al.* 2005 [8] |
| CaIrO$_3$ | 19.4 | 4.32 | McDaniel *et al.* 1972 [43]/Hirai *et al.* 2009 [42] |
| CaPtO$_3$ | no perovskite known | 4.46 | Ohgushi *et al.* 2008 [3] |
| CaRuO$_3$ | 16.2 | 4.25 | Kojitani *et al.* 2007 [5] |
| CaRhO$_3$ | 19.2 / 19.0 | 4.41 | Yamaura *et al.* 2009 [34]/Shirako *et al.* 2009 [6] |
| NaMgF$_3$ | 15.2 | 3.94 | Martin *et al.* 2006 [41] |

* Values calculated using the reported structural parameters under pressure because these post-perovskites cannot be recovered to ambient conditions, 121 GPa for MgSiO$_3$ and 78 GPa for MgGeO$_3$.

**Table 5. Goldsmith tolerance factor for perovskite ($t$), post-perovskite tolerance factor ($t'$), and electronegativity difference ($\Delta\chi_P$) for ABX$_3$-type post-perovskite compounds. Ionic radii are from Shannon[35] with coordination number (CN): CN(A) = 8, CN(B) = 6, CN(X) = 6 except for U$^{3+}$, where CN = 6.**

| Compound | Goldsmith $t$ | Post-perovskite $t'$ | $\Delta\chi_P$ | Reference |
|---|---|---|---|---|
| NaIrO$_3$ | 0.926 | 0.958 | 1.24 | *present study* |
| MgSiO$_3$ | 0.900 | 0.837 | 1.54 | Murakami *et al.* 2004 [1] |
| MgGeO$_3$ | 0.839 | 0.865 | 1.43 | Hirose *et al.* 2005 [8] |
| CaIrO$_3$ | 0.880 | 0.854 | 1.24 | Hirai *et al.* 2009 [42] |
| CaPtO$_3$ | 0.880 | 0.872 | 1.16 | Ohgushi *et al.* 2008 [3] |
| CaRuO$_3$ | 0.882 | 0.858 | 1.24 | Kojitani *et al.* 2007 [5] |
| CaRhO$_3$ | 0.891 | 0.870 | 1.16 | Shirako *et al.* 2009 [6] |
| NaMgF$_3$ | 0.866 | 0.845 | 2.67 | Martin *et al.* 2006 [41] |
| KTmI$_3$ | 0.812 | 0.828 | 1.41 | Schilling *et al.* 1992 [44] |
| UScS$_3$ | 0.784 | 0.806 | 1.22 | Julien *et al.* 1978 [45] |
| ThMnSe$_3$ | 0.765 | 0.789 | 1.00 | Ijjaali *et al.* 2004 [12] |
| UMnSe$_3$ | 0.752 | 0.779 | 1.00 | Ijjaali *et al.* 2004 [12] |
| LaYbS$_3$ | 0.783 | 0.799 | 1.48 | Mitchell *et al.* 2004 [46] |
| *RE*YbSe$_3$* | 0.759-0.780 | 0.780-0.797 | 1.45 | Mitchell *et al.* 2004 [46] |

* *RE* = La, Ce, Pr, Nd, Sm.

**Figure Captions**

**Fig. 1.** A) Rietveld refinement of the powder X-ray diffraction pattern of NaIrO$_3$ ($\lambda$ = 0.69910 Å). The second row of Bragg reflections indicate impurity peaks from metallic iridium, which in this sample occurs at the 20 wt% level. The inset shows the unit cell of the post-perovskite NaIrO$_3$. B) Edge-sharing IrO$_6$ octahedra viewed along the *b*-axis, and C) viewed along the corner-sharing oxygen to iridium bond. D) The NaO$_8$ bicapped trigonal prism.

**Fig. 2.** The magnetic susceptibility, $\chi$, of NaIrO$_3$. The inset shows a Curie-Weiss plot with a fit to the Curie-Weiss law, Eq. 1.

**Fig. 3.** Resistivity ($\rho$) of NaIrO$_3$ measured from 2 to 300 K showing highly resistive behavior. Inset A shows log($\rho$) vs. $1/T$ in the temperature range corresponding to $50 < T < 300$ K. Inset B shows a fit to a variable range hopping model with dimension, $d = 2$, in the temperature range $2 < T < 50$ K.

**Fig. 4.** $C_P/T$ vs. $T^2$ at low temperatures. The solid blue line is a fit to Eq. 2 and the dotted red line is a fit to $C_p$ for $3\ \text{K} < T < 10\ \text{K}$ excluding the *A*-term. The inset shows the specific heat, $C_p$ vs. $T$ from 0.5 to 20 K.

**Fig. 5.** Electronic band structure of NaIrO$_3$ calculated with spin-polarization LDA+$U$+SO. B) The corresponding DOS for NaIrO$_3$ as shown in A. C) The DOS obtained from a simple LDA calculation for comparison.

*Fig. 1*

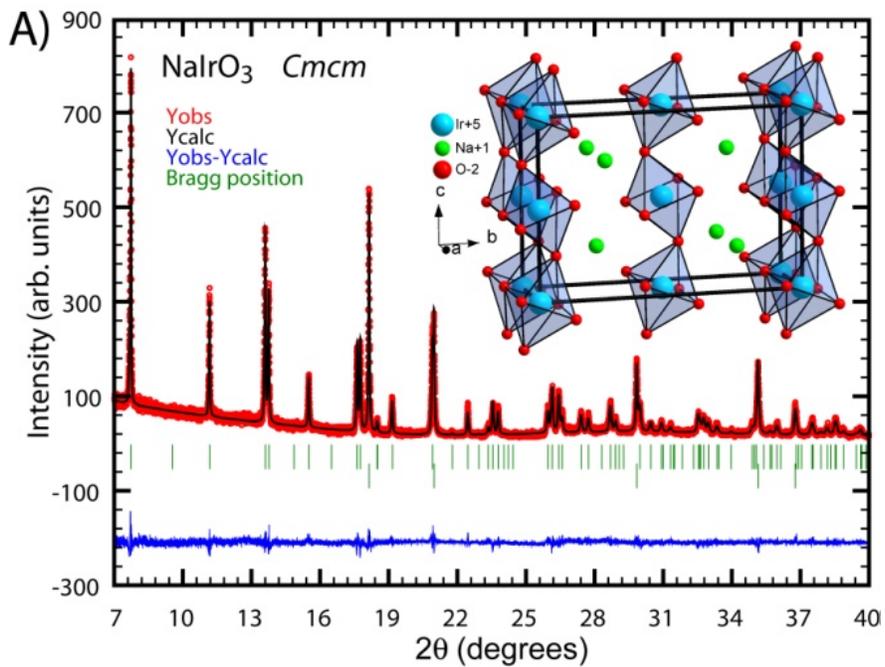

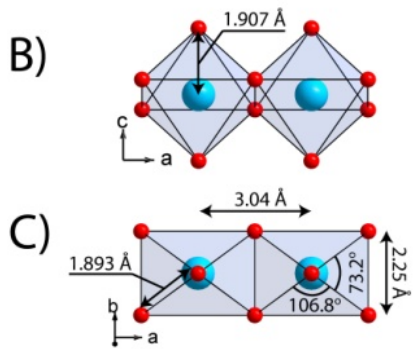

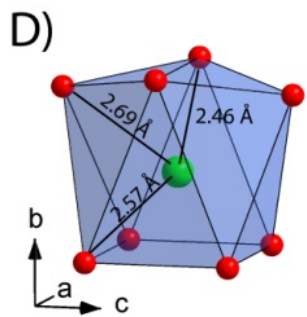

*Fig. 2*

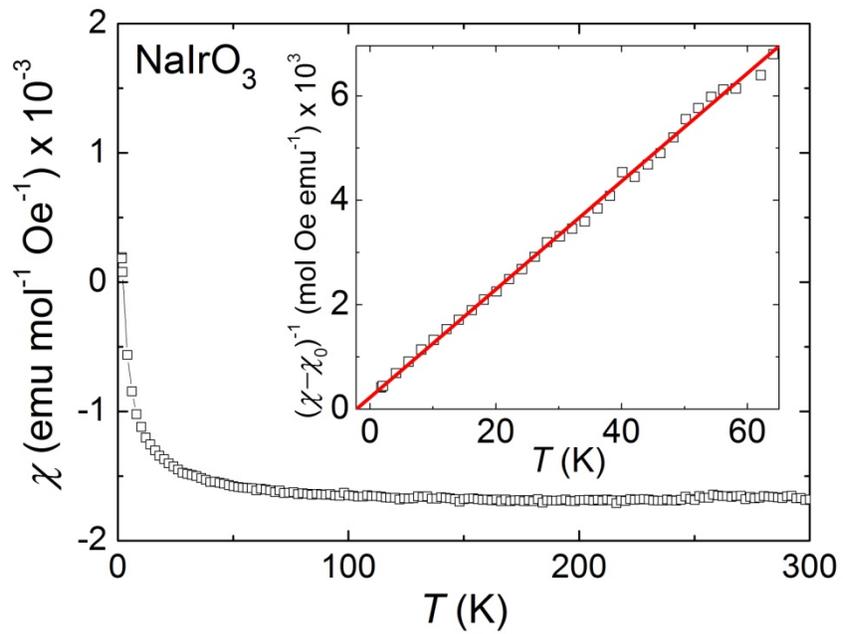

*Fig. 3*

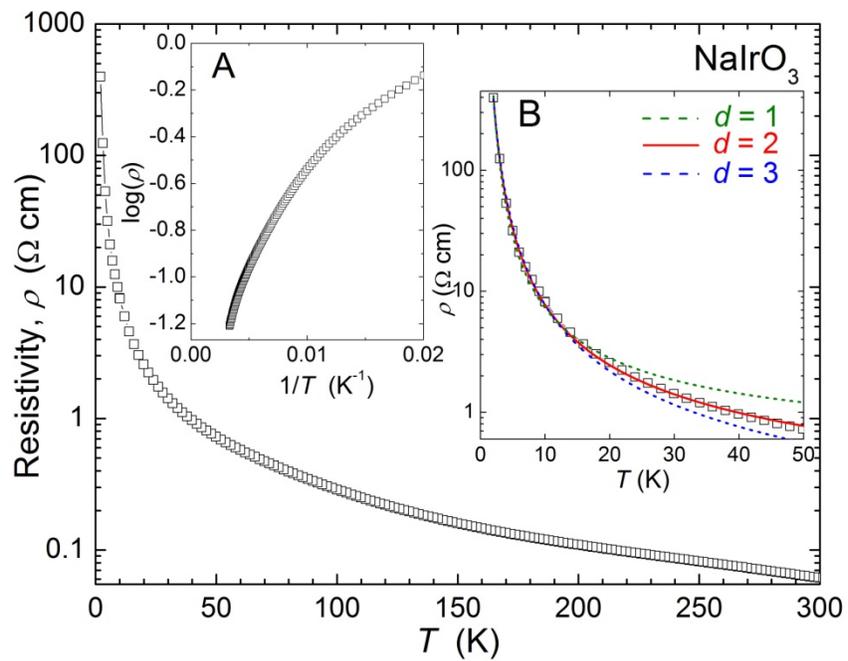

*Fig. 4*

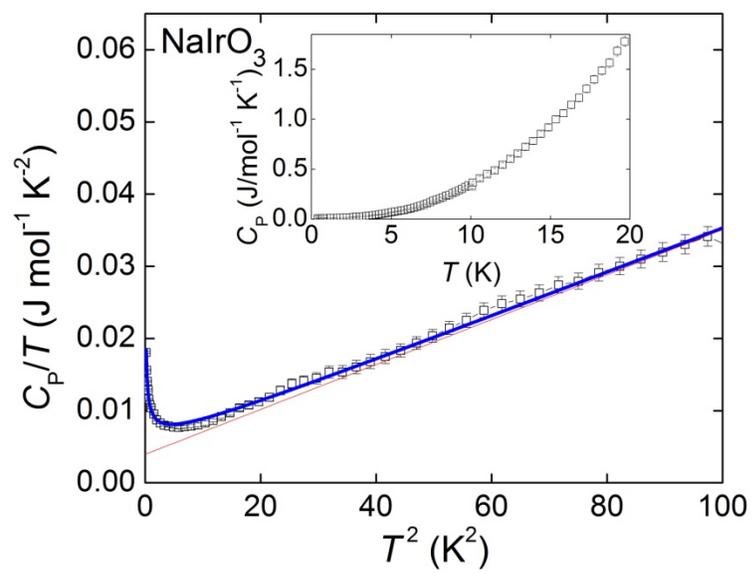

*Fig. 5*

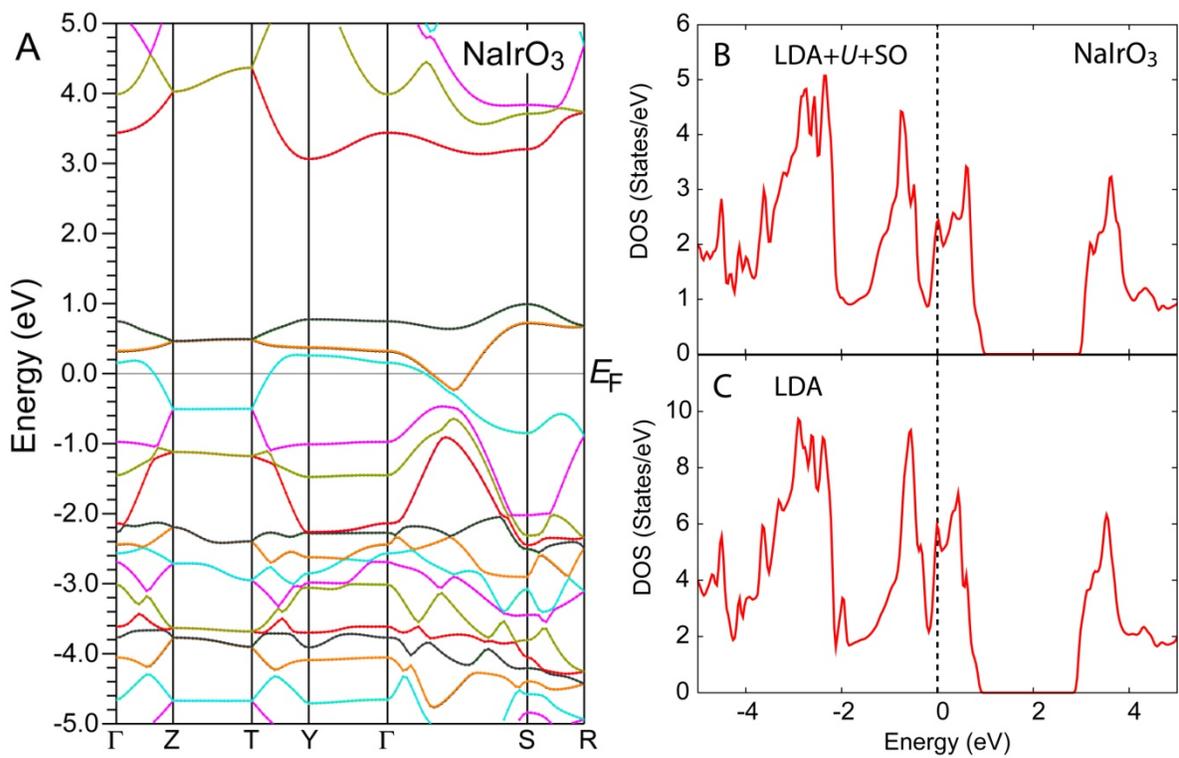